\documentclass[aps,prl,twocolumn,showpacs]{revtex4}

\usepackage{graphicx}
\begin{document}

%\preprint{}

\title{Quantum criticality and disorder in the antiferromagnetic critical point of NiS$_{2}$ pyrite}

\author{N. Takeshita$^{1}$, S. Takashima$^{2}$, C. Terakura$^{1}$, H. Nishikubo$^{2}$, S. Miyasaka$^{3}$, M. Nohara$^{2}$, Y. Tokura$^{1,4}$, and H. Takagi$^{1,2}$}
\affiliation{
$^{1}$Correlated Electron Research Center (CERC), National Institute of Advanced Industrial Science and Technology (AIST), Tsukuba 305-8562, Japan\\
$^{2}$Department of Advanced Materials, University of Tokyo, Kashiwa 277-8561, Japan\\
$^{3}$Department of Physics, Osaka University, Toyonaka 560-0043, Japan\\
$^{4}$Department of Applied Physics, University of Tokyo, Tokyo 113-8656, Japan}

\date{\today}

\begin{abstract}
A quantum critical point (QCP) between the antiferromagnetic and the paramagnetic phases was realized by applying a hydrostatic pressure of ${\sim}$ 7 GPa on single crystals of NiS$_{2}$ pyrite with a low residual resistivity, $\rho_{0}$, of 0.5 $\mu\Omega$cm. We found that the critical behavior of the resistivity, $\rho$, in this clean system contrasts sharply with those observed in its disordered analogue, NiS$_{2-x}$Se$_{x}$ solid-solution, demonstrating the unexpectedly drastic effect of disorder on the quantum criticality. Over a whole paramagnetic region investigated up to $P$ = 9 GPa, a crossover temperature, defined as the onset of T$^{2}$ dependence of $\rho$, an indication of Fermi liquid, was suppressed to a substantially low temperature 
$T$ $\sim$ 2 K and, instead, a non Fermi liquid behavior of $\rho$, $T^{3/2}$-dependence, robustly showed up. 
%No trace of superconductivity was found down to 180 mK at around ${\sim}$ 7 GPa, suggesting a need for additional ingredient to induce superconductivity at QCP.
\end{abstract}

\maketitle

A hallmark of strongly correlated electron systems is the presence of a rich variety of phases often competing with each other. When two phases meet with each other in the $T$ = 0 limit by tuning a phase controlling parameter such as pressure and chemical doping, quantum fluctuations often give rise to exotic states of electrons, which has been attracting considerable interest in condensed matter research \cite{ref1}. One of the most fascinating cases is a breakdown of the Fermi liquid at magnetic quantum critical points (QCP) in itinerant magnets, which has been believed to be captured by self-consistent renormalization theory \cite{ref2} and scaling analysis \cite{ref3,ref4}. The onset temperature of Fermi liquid coherence, probed by a quadratic temperature dependence of resistivity $\rho(T)$ $\propto$ $T^{2}$, is predicted to be suppressed by the presence of low lying spin fluctuations near QCP and, right at QCP, a non Fermi liquid ground state is realized which manifests itself as a non-trivial power law behavior of resistivity $\rho(T)$ $\propto$ $T^{n}$ down to the $T$ = 0 limit, where $n$ = 3/2 for antiferromagnetic (AF) QCP and $n$ = 5/3 for ferromagnetic (F) QCP \cite{ref5}. A V-shaped recovery of Fermi liquid behavior, $T^{2}$-resistivity, around QCP is anticipated as a function of phase tuning parameter.

The critical behavior of $\rho(T)$ near the AF QCP in NiS$_{2-x}$Se$_{x}$ solid 
solution is a textbook demonstration of standard theories for QCP. 
NiS$_{2}$ crystallizes in the pyrite structure. Ni is divalent and therefore accommodates two electrons in doubly degenerate $e_{g}$ orbitals ($t_{2g}^{6}e_{g}^{2}$). 
Due to a strong onsite Coulomb repulsion among Ni $e_{g}$ electrons, the system is a $S$ = 1 Mott insulator \cite{ref13,ref14}. By substituting S with Se, 
the effective bandwidth can be increased due to the increase of $p$-$d$ hybridization. With increasing $x$ in NiS$_{2-x}$Se$_{x}$, the system experiences a weakly first order transition to an AF metal with a collinear spin structure \cite{ref15} at $x$ ${\sim}$ 0.4 and then a second order transition to a paramagnetic metal at $x$ = 1.0.

In the AF metal phase of NiS$_{2-x}$Se$_{x}$, the AF transition temperature is $T_{N}$ $\sim$ 90 K at $x$ = 0.5 and, with increasing $x$, continuously goes down to $T$ = 0 at $x$ = 1.0, giving rise to a well defined AF QCP. 
The $T^{3/2}$ dependence of $\rho(T)$, expected for AF QCP, is observed at least down to 1.7 K at $x$ = 1.0. On going away from $x$ = 1.0, $T^{2}$-behavior of $\rho(T)$ quickly recovers and a V-shaped region with $T^{2}$ resistivity is identified around $x$ = 1.0. It is known that the application of pressure is equivalent to Se substitution in that it increases the band width. 
By applying pressure $P$ on an AF metal NiS$_{1.3}$Se$_{0.7}$, suppression of $T_{N}$ analogous to Se substitiution was indeed observed and, eventually, QCP was approached with $P$ = 2 GPa \cite{ref16}. 
The phase diagram and the critical behavior of the resistivity in pressurized NiS$_{1.3}$Se$_{0.7}$ were essentially identical with Se content $x$ simply replaced with $P$ \cite{ref16}.

Recently, however, there has been growing evidence that, the above mentioned textbook picture is violated in a variety of intermetallic systems. In a helical magnet MnSi \cite{ref6} and a weak ferromagnet ZrZn$_{2}$ \cite{ref7}, 
a non trivial power law behavior of resistivity, $\rho(T)$ $\propto$ $T^{3/2}$, dominates the resistivity down to a very low temperature, not only right at the QCP but also over a wide range of paramagnetic phase. At the AF QCP in CePd$_{2}$Si$_{2}$, with increasing the purity of sample, the exponent of the power law resistivity was found to deviate significantly from the standard value of 3/2 \cite{ref8}. 
The common feature among these systems is that they are clean with a low 
residual resistivity of $\rho_{0}$ $<$ 1 $\mu\Omega$cm \cite{ref6,ref7}. 
In contrast, in NiS$_{2-x}$Se$_{x}$ solid solution where textbook example of critical behavior of $\rho(T)$ is observed, Se substitution inherently gives rise to disorder. Indeed, $\rho_{0}$ of NiS$_{2-x}$Se$_{x}$ around QCP is as large as several 10 $\mu\Omega$cm. Questions immediately arise. Does the non-trvial behavior in the intermetallics represent a generic property of magnetic QCP in clean systems? Does standard behavior of QCP shows up only when the system is disordered? To tackle these questions experimentally, we attempted to realize a ``clean" QCP in NiS$_{2-x}$Se$_{x}$. The parent compound of NiS$_{2-x}$Se$_{x}$, NiS$_{2}$, is pure and presumably clean. 
If one can approach the QCP of pure NiS$_{2}$ by pressure without relying on Se substitution, a clean analogue of AF QCP in NiS$_{2-x}$Se$_{x}$ can be explored and the impact of disorder on QCP can be captured. Recent progress of high pressure technique enabled us to do so.

In this Letter, we address the issue of criticality and disorder by examining 
the critical behavior of resistivity of pure NiS$_{2}$ under pressures. 
The AF QCP of NiS$_{2}$ was reached at $\sim$ 7 GPa, where the system was found very clean with a low residual resistivity $\rho_{0}$ of $\sim$ 0.5 $\mu\Omega$cm. Not only right at the QCP but over an entire range of the paramagnetic phase investigated, the recovery of Fermi liquid $T^{2}$ of $\rho(T)$ is suppressed substantially to a very low temperature below $\sim$ 2 K and non Fermi liquid behavior with $T^{3/2}$ dependence of $\rho(T)$ dominated. 
We demonstrate the drastic influence of disorder on this AF QCP by contrasting 
the result with previous pressure work on NiS$_{1.3}$Se$_{0.7}$ 
with a residual resistivity of 60 $\mu\Omega$cm \cite{ref16}.

NiS$_{2}$ sample used in this study was prepared by a vapor transport technique. 
The resistivity measurement was performed by a conventional four probe technique 
under hydrostatic pressure up to $\sim$ 10 GPa in a cubic anvil type 
pressure system down to 3 K and also in a modified Bridgman-type pressure cell 
down to 180 mK. The results obtained by the two different pressure setups 
agreed reasonably in the temperature range of overlap, indicating a very good 
homogeneity of pressure. Pressure was calibrated by measuring the superconducting transition temperature of Pb \cite{ref17}.

The inset of Fig. 1 demonstrates $\rho(T)$ of NiS$_{2}$ at relatively low pressures below 4 GPa. With applying pressure, the insulating behavior of $\rho(T)$ switches into metallic behavior, indicating the occurrence of metal-insulator transition. In between 2.6-3.4 GPa, we observe a discontinuous jump of resistivity as a function of temperature, which corresponds to a first order metal-insulator transition line on the phase diagram in Fig. 1. The discontinuous jump appears to terminate around 200 K, indicating the presence of a critical end point. In the phase diagram of NiS$_{2-x}$Se$_{x}$ solid solution, the first order phase line terminates at much lower temperature and is hard to identify \cite{ref18}. This difference appears to suggest the strong influence of disorder on the Mott criticality.

\begin{figure}[t]
\includegraphics[width=0.95\linewidth]{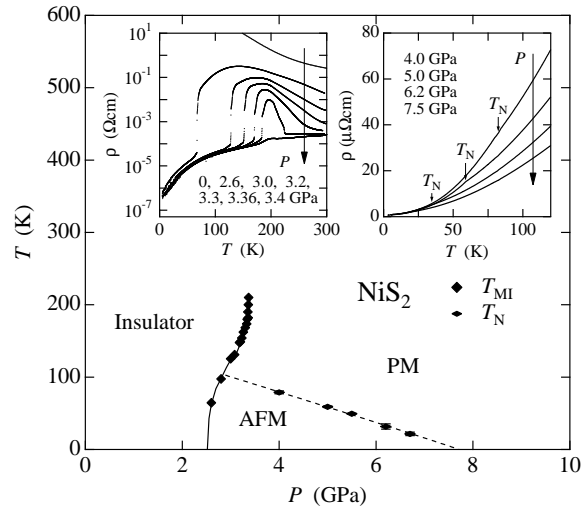}
\caption{The electronic phase diagram of clean NiS$_{2}$ pyrite as a function of pressure. PM and AFM denote paramagnetic metal and AF metal, respectively. The inset shows the temperature dependent resistivity under pressures, $P$ = 0 - 3.4 GPa (left) and $P$ = 4.0 - 7.5 GPa (right).}
\label{fig1}
\end{figure}

As seen in the inset of Fig. 1, $\rho(T)$ of pure NiS$_{2}$ showed metallic behavior above $P$ = 2.6 GPa. The residual resistivity at the critical point was as low as $\sim$ 0.5 $\mu\Omega$cm, demonstrating that the system is indeed very clean. Magnetic ordering in the AF metal phase manifests itself as a very weak but sharp kink in $\rho(T)$ at $T_{N}$ as indicated by the arrows. 
The antiferromagnetic transition temperature $T_{N}$ thus determined systematically goes down upon pressure and approaches $T$ = 0 somewhere around 7-7.5 GPa. 
No superconductivity was observed between $P$ = 6 and 9.1 GPa down to 180 mK, 
in spite of the low residual resistivity. This appears to suggest that realizing an AF QCP in clean systems alone is not enough to achieve exotic superconductivity as observed in heavy Fermion compounds \cite{ref9,ref10,ref11,ref12} and that additional ingredients such as Kondo physics must be invoked.

The pressure dependence of $T_{N}$, determined by the kink in $\rho(T)$, 
together with the first order metal insulator transition, is summarized as a phase diagram in Fig. 1. $T_{N}$ appears to decrease almost linearly in contradiction to $(P_{c}-P)^{2/3}$ dependence expected from self consistent renormalization theory \cite{ref5}. Unusual linear suppression of the magnetic transition temperature was also observed analogously for a helical magnet MnSi \cite{ref6} and a weakly ferromagnet ZrZn$_{2}$ \cite{ref7} when the sample is very clean. It may be interesting to infer that, in these clean system, the magnetic transition as a function of pressure is reported to be a first order rather than a second order. 
In the clean NiS$_{2}$, we cannot rule out the possibility of a first order 
transition at this stage, because $\rho(T)$ is not very sensitive to $T_{N}$ near the critical point.

\begin{figure}[t]
\includegraphics[width=0.85\linewidth]{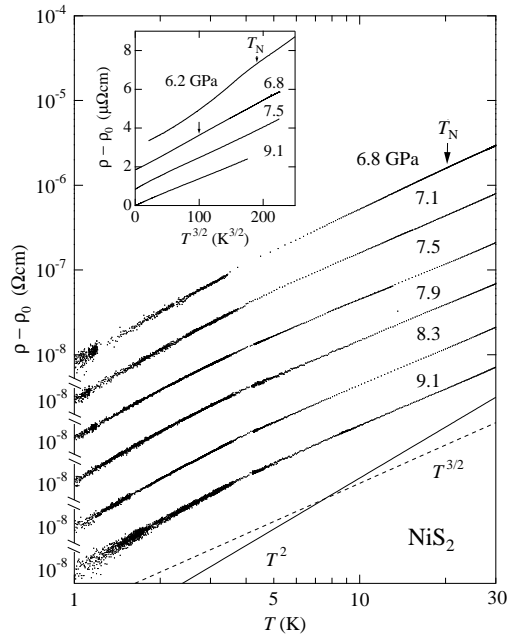}
\caption{Temperature dependent part of resistivity $\rho - \rho_{0}$ as a function of temperature under pressure above $\sim$ 7 GPa, where the system is paramagnetic, plotted as $\log(\rho-\rho_{0})$ vs. $\log T$. The inset shows $\rho$ vs. $T^{3/2}$ plot.}
\label{fig2}
\end{figure}

The signature of AF criticality in this clean system was explored. 
The inset of Fig. 2 demonstrates $\rho(T)$ below 30 K, plotted as $\rho$ vs. $T^{3/2}$. In the antiferromagnetic phase at $P$ = 6.2 GPa, $\rho$ - $T^{3/2}$ curve is linear above $T_{N}$ but shows superlinear behavior below $T_{N}$. The temperature dependence below $T_{N}$ is found to be approximately $T^{2}$, indicative of the formation of coherent quasi particles. 
In the paramagnetic phase above $\sim$ 7 GPa, however, the $\rho$ - $T^{3/2}$ 
curve shows a linear behavior down to very low temperature which is expected 
for the antiferromagnetic critical point due to low lying spin fluctuations. 
It is remarkable to see $T^{3/2}$ behavior characteristic of the antiferromagnetic critical point over such a wide range of pressure from $\sim$ 7 GPa to $\sim$ 9 GPa. 
$\rho(T)$ is surprisingly insensitive to the pressure in the paramagnetic region above 7 GPa and it is hard to find a signature of criticality. This is analogous to those observed in a helical magnet MnSi \cite{ref6} and a weakly ferromagnetic magnet ZrZn$_{2}$ \cite{ref7} when the sample is clean, implying that unusual critical behavior in clean systems is ubiquitous.

\begin{figure}[t]
\includegraphics[width=0.95\linewidth]{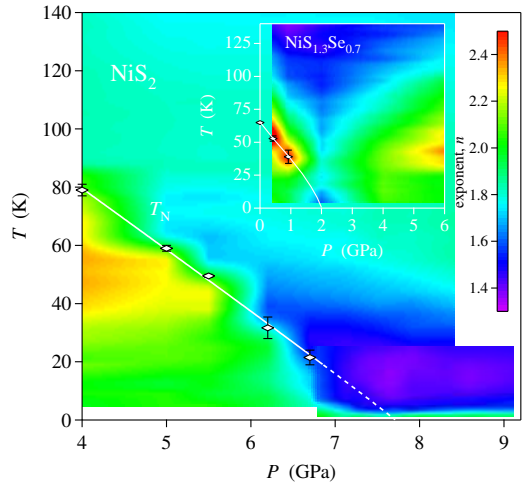}
\caption{Contour plot of the exponent $n$ of power low dependence of resistivity on pressure-temperature plane, demonstrating the criticality observed in the temperature dependence of resistivity. 
The main panel is data for clean NiS$_{2}$ and the inset shows data for dirty NiS$_{2-x}$Se$_{x}$. The N${\rm\acute{e}}$el temperature determined by resistivity anomaly was shown by the white line.}
\label{fig3}
\end{figure}

To investigate the details of unusual temperature dependence in the paramagnetic phase further, we plotted the temperature dependence of $\rho-\rho_{0}$ as $\log(\rho-\rho_{0})$ vs. $\log T$ in the main panel of Fig. 2. The residual resistivity $\rho_{0}$ was determined by extrapolating $\rho$ - $T^{2}$ curve to $T$ = 0 limit. We note here that the temperature dependent part $\rho-\rho_{0}$ is comparable to $\rho_{0}$ at $\sim$ 3 K and, therefore, the ambiguity originating from the estimate of $\rho_{0}$ does not influence the temperature dependence of $\rho-\rho_{0}$ at least above 1 K. It is again clear that the slope is apparently smaller than 2 and instead close to 3/2 above $\sim$ 2 K. At the lowest temperatures below $\sim$ 2 K, however, the slope becomes steeper and $T^{2}$-resitivity appears to recover eventually below 2 K. This crossover temperature to $T^{2}$-resistivity is again insensitive to pressure and always 2-3 K up to 9 GPa. Close inspection of data indicates that the crossover temperature is the lowest around 7.5 GPa but only slightly lower than the other pressures.

This strong suppression of the crossover to $T^{2}$ behavior in $\rho(T)$, 
over a remarkably wide range of pressure from $\sim$ 7 GPa up to $\sim$ 9 GPa, 
contrasts sharply with the observation in Se-doped samples, where the 
recovery of $T^{2}$ resistivity was clearly observed not only for 
the magnetic side but also for the paramagnetic side. As a function of 
Se-doping, a metal-insulator transition and antiferromagnetic critical point 
occurs at around Se content $x$ = 0.4 and 1.0, respectively, while $\sim$ 2.5 GPa and $\sim$ 7 GPa are required to reach a metal-insulator transition 
and QCP, respectively. This yields a conversion ratio of phase controlling parameters, $\sim$ 0.15 Se/1 GPa. In this disordered NiS$_{2-x}$Se$_{x}$, 
the $T^{3/2}$-dependence of $\rho(T)$ dominates at the QCP of $x$ = 1.0. With further doping of Se up to $x$ = 1.33 which is equivalent of additional pressure of $\simeq$ 2 GPa, however, the $T^{2}$ resistivity is fully recovered and can be observed below $\sim$ 80 K \cite{ref16}. 
Analogously, in a Se doped NiS$_{1.3}$Se$_{0.7}$ crystal under pressure, 
on going from the QCP at $P$ $\sim$ 2 GPa to $P$ = 4 GPa, $T^{2}$ resistivity 
recovers quickly and shows up below 80 K with increase of 2 GPa. These should be compared with the low crossover temperature of 2-3 K, approximately 2 GPa above the critical point. 

To visually illustrate these points, we plotted the exponent of power law dependence of $\rho(T)$, $n$ as a contour map on the pressure-temperature plane in Fig. 3. The exponent was determined by taking the derivative of the 
$\log(\rho-\rho_{0})$ - $\log T$ curve in Fig. 2. It is clear that the 
V-shaped recovery of Fermi liquid behavior around QCP is absent in clean NiS$_{2}$. The recovery can be seen only on the antiferromagnetic side below 7 GPa, where the region with $n$ = 2 ($T^{2}$) develops below $T_{N}$. Above the critical point of $P$ $\sim$ 7 GPa, it is clear that the $n$ = 1.5 ($T^{3/2}$) region predominantly occupies a majority of the paramagnetic phase. A thin region with a different color is lying at the $T$ = 0 limit. This corresponds to the marginal recovery of Fermi liquid behavior below $\sim$ 2 K. In the inset of Fig. 3, we have constructed the contour map also for the NiS$_{1.3}$Se$_{0.7}$ data under pressure from a previous report \cite{ref14}. Note again the V-shaped recovery on the temperature scale of 100 K over $\sim$ 2 GPa pressure.

The remarkable contrast in the critical behavior between pure NiS$_{2}$ 
and NiS$_{2-x}$Se$_{x}$, visually demonstrated in Fig. 3, indicates that 
the influence of disorder on quantum criticality is surprisingly drastic, 
since the only difference between the two systems is the disorder 
produced by Se substitution. In NiS$_{1.3}$Se$_{0.7}$ solid solution, 
the residual resistivity $\rho_{0}$ is approximately 60 $\mu\Omega$cm, 
which is larger than those of pure NiS$_{2}$ by two orders of magnitude. 
When the samples are disordered, we do see a canonical behavior of the QCP as 
predicted by standard theories \cite{ref3,ref4,ref5}. 
To our surprise, once the system becomes clean, the textbook behavior 
is gone and the Fermi liquid coherence seen in $\rho(T)$ is dramatically suppressed. We should note that the magnitude of $\rho(T)-\rho_{0}$ is roughly 50 ${\mu\Omega}$cm in the temperature range below 100 K at around QCP. In the Se doped crystal, inelastic scattering is always weaker than or at most comparable to elastic scattering due to disorder below 100 K. 
In the pure NiS$_{2}$, in contrast, the same situation, $\rho-\rho_{0}$ $<$ $\rho_{0}$ occurs only below 2-3 K, where we observed crossover to $T^{2}$-resistivity. This suggests that disorder might be controlling the appearance of $T^{2}$-resistivity.

One of the plausible scenarios for the strong influence of disorder and robust 
non-Fermi liquid behavior might be a dichotomy of the Fermi surface \cite{ref19}. It is natural that a specific part of Fermi surface, a ``hot spot", is coupled strongly with a critical antiferromagnetic fluctuation with a characteristic wave vector $Q$. 
There may remains a region with well defined quasiparticles free from critical fluctuations, a cold spot. 
The transport is then determined by an interplay of these two contributions 
at high temperatures but eventually a cold spot with $T^{2}$-dependence should dominate the conduction at very low temperatures. 
This phase separation in $k$-space might explain in part the unusual 
temperature dependence observed in pure NiS$_{2}$ but it is not clear 
whether the robustness of non Fermi liquid behavior can be properly described. 
In this scenario, the strong influence of disorder can be naturally understood. 
The strong elastic scatting should mix up the hot spot and cold spot 
and the inelastic scattering therefore becomes effectively isotropic, 
which might be close to the situation implicitly assumed in standard theories \cite{ref5}. 
Another scenario might be a phase separation and the resultant domain or bubble formation in real space as discussed in clean MnSi where the helical spin order disappears discontinuously as a first order transition \cite{ref6}. 
These bubbles and domains have been proposed to be responsible for 
the robust non Fermi liquid behavior in the paramagnetic phase. 
It is worth checking the possible first order transition carefully checking the magnetism at $\sim$ 7 GPa.

In conclusion, we have demonstrated the sharp contrast in the quantum critical behavior of $\rho(T)$ between the clean and the disordered systems by examining a single crystal of NiS$_{2}$ with a low residual resistivity of $\sim$ 0.5 $\mu\Omega$cm. Previously, the V-shaped recovery of Fermi liquid behavior ($T^{2}$-behavior of resistivity) around the antiferromagnetic critical point was clearly 
observed as a function of pressure and Se content in the dirty NiS$_{2-x}$Se$_{x}$ systems with $\rho-\rho_{0}$ $<$ $\rho_{0}$. In sharp contrast, we found a robust non Fermi liquid behavior over a wide pressure range in the paramagnetic side of a QCP in the clean system with $\rho-\rho_{0}$ $\gg$ $\rho_{0}$. 
These results clearly demonstrate that our understanding of the quantum critical point is still far from complete and some important ingredient must be missing.

The authors would like to thank M. Imada and H. Fukuyama for discussion. 
This work was partly supported by a Grant-in-Aid for Scientific Research (No. 18043009) from the Ministry of Education, Culture, Sports, Science and Technology of Japan and CREST, Japan Science and Technology Agency (JST).

\end{document}